\documentclass[
 reprint,
 superscriptaddress,
 preprintnumbers,
 amsmath
 amssymb,
 prx,
 floatfix,
]{revtex4-2}

\DeclareMathAlphabet{\mathsfit}{T1}{\sfdefault}{\mddefault}{\sldefault}
\SetMathAlphabet{\mathsfit}{bold}{T1}{\sfdefault}{\bfdefault}{\sldefault}

\usepackage{amsmath}
\usepackage{graphicx}
\graphicspath{{fig/}}   
\usepackage{dcolumn}
\usepackage{bm}
\usepackage{makecell}
\usepackage{braket} 
\usepackage{siunitx}
\usepackage{color}
\usepackage{array}
\usepackage{xr}
\usepackage{upgreek}
\usepackage[colorlinks]{hyperref}
\usepackage[capitalize]{cleveref}
\usepackage[caption=false]{subfig}
\usepackage{multirow} 
\usepackage[version=4]{mhchem} 

\usepackage{pdfpages} 
\usepackage{pgffor} 
\makeatletter
\AtBeginDocument{\let\LS@rot\@undefined}
\makeatother
\def\supplementfilename{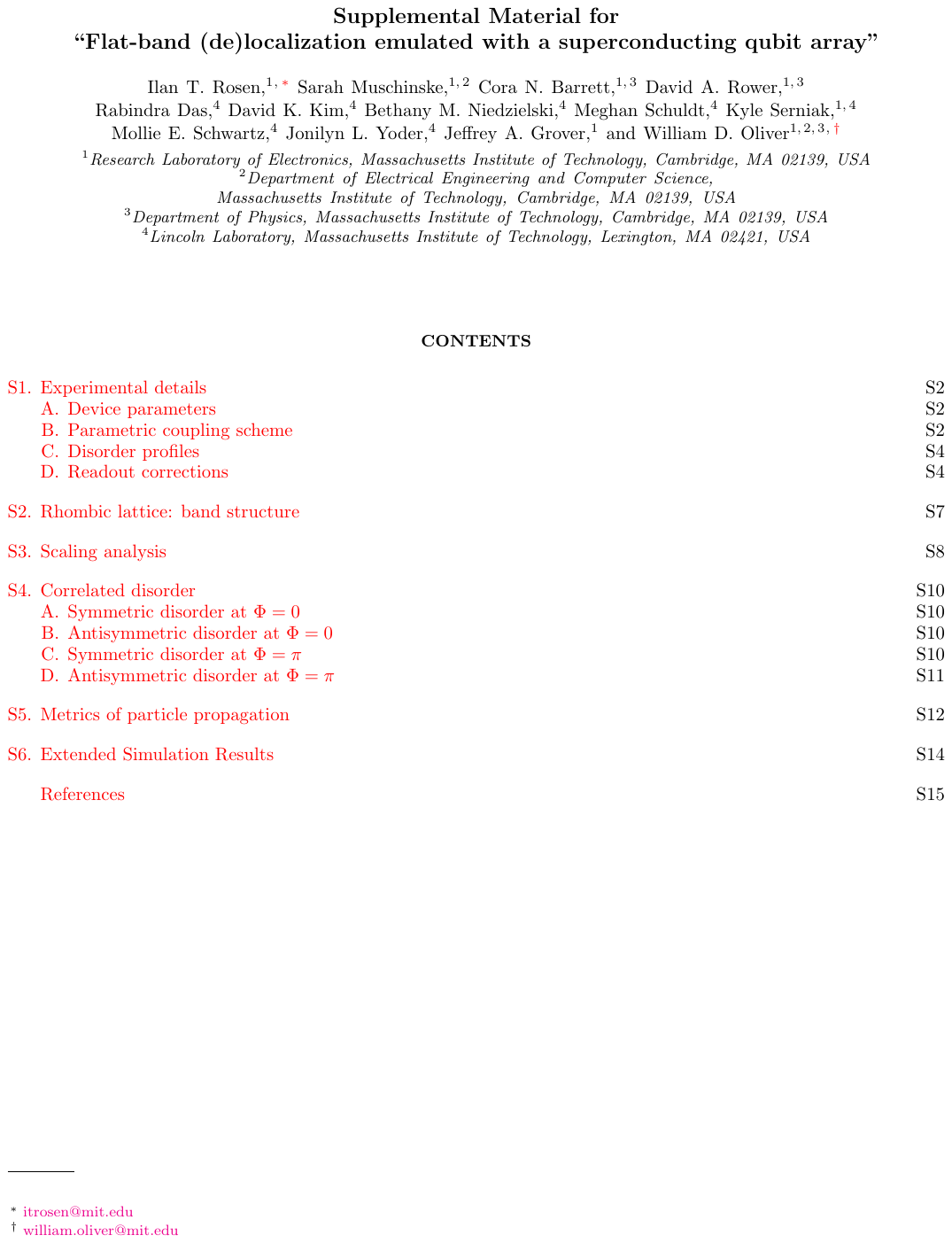}
\pdfximage{\supplementfilename}
\def\numbersupplementpages{\the\pdflastximagepages}

\newcommand{\supp}{Supplemental Material}

\begin{document} 

\title{Flat-band (de)localization emulated with a superconducting qubit array}

\def\RLEaffil{Research Laboratory of Electronics, Massachusetts Institute of Technology, Cambridge, MA 02139, USA}
\def\LLaffil{Lincoln Laboratory, Massachusetts Institute of Technology, Lexington, MA 02421, USA}
\def\Physaffil{Department of Physics, Massachusetts Institute of Technology, Cambridge, MA 02139, USA}
\def\EECSaffil{Department of Electrical Engineering and Computer Science, Massachusetts Institute of Technology, Cambridge, MA 02139, USA}

\author{Ilan~T.~Rosen}
\email{itrosen@mit.edu}
\affiliation{\RLEaffil}

\author{Sarah~Muschinske}
\affiliation{\RLEaffil}
\affiliation{\EECSaffil}

\author{Cora~N.~Barrett}
\affiliation{\RLEaffil} 
\affiliation{\Physaffil}

\author{David~A.~Rower}
\affiliation{\RLEaffil} 
\affiliation{\Physaffil}

\author{Rabindra~Das}
\affiliation{\LLaffil}

\author{David~K.~Kim}
\affiliation{\LLaffil}

\author{Bethany~M.~Niedzielski}
\affiliation{\LLaffil}

\author{Meghan~Schuldt}
\affiliation{\LLaffil}

\author{Kyle~Serniak}
\affiliation{\RLEaffil}
\affiliation{\LLaffil}

\author{Mollie~E.~Schwartz}
\affiliation{\LLaffil}

\author{Jonilyn~L.~Yoder}
\affiliation{\LLaffil}

\author{Jeffrey~A.~Grover}
\affiliation{\RLEaffil}

\author{William~D.~Oliver}
\email{william.oliver@mit.edu}
\affiliation{\RLEaffil}
\affiliation{\EECSaffil}
\affiliation{\Physaffil}

\begin{abstract}
Arrays of coupled superconducting qubits are analog quantum simulators able to emulate a wide range of tight-binding models in parameter regimes that are difficult to access or adjust in natural materials.
In this work, we use a superconducting qubit array to emulate a tight-binding model on the rhombic lattice, which features flat bands.
Enabled by broad adjustability of the dispersion of the energy bands and of on-site disorder, we examine regimes where flat-band localization and Anderson localization compete.
We observe disorder-induced localization for dispersive bands and disorder-induced delocalization for flat bands.
Remarkably, we find a sudden transition between the two regimes and, in its vicinity, the semblance of quantum critical scaling.
\end{abstract}

\maketitle


\section{Introduction}

Several mechanisms arrest the propagation of particles through partially-filled lattices of coupled sites.
First, disorder in the lattice, particularly variation of the self-energy of each site, leads to an Anderson insulator~\cite{anderson1958, abrahams1979}.
Second, interactions between particles can lead to a Mott insulator~\cite{mott1968}.
Third, quantum interference suppresses particle propagation in lattices with flat-band structures~\cite{sutherland1986}, an effect related to geometric frustration~\cite{leykam2018}.
Additionally, dc and ac electric fields can cause Wannier-Stark~\cite{emin1987} and dynamical localization~\cite{dunlap1986}, respectively.
As these effects may underpin metal-insulator transitions in a variety of materials, it is natural to consider the realistic scenario where multiple mechanisms are simultaneously relevant.
For example, a body of literature has emerged studying the Mott-Anderson regime, where disorder and interactions compete~\cite{belitz1994, aguiar2009}.

The transition between an Anderson insulator and a flat-band insulator, however, is relatively unexplored.
Interest in flat-band systems has surged following recent realizations in van der Waals heterostructures and kagome-structured crystals~\cite{cao2018-mott, cao2018-sc, chen2019, sharpe2019, kang2020, zeng2023, cai2023, wakefield2023, checkelsky2024}.
Flat bands have quenched kinetic energy, therefore the effects of disorder and interactions are magnified, leading to a rich set of quantum phases found in these materials.
Yet, whereas Mott physics in two-dimensional materials may readily be tuned by adjusting carrier density through electrostatic gating~\cite{cao2018-mott, chen2019}, disorder and band flatness are both difficult to adjust in natural materials.

An alternative experimental approach is to emulate lattice models in artificial materials
and to tune parameters throughout regimes of interest.
In particular, arrays of coupled superconducting qubits natively emulate the Bose-Hubbard model while offering programmable on-site energy and inter-site coupling strength, so that disorder and band structure may be continuously tuned~\cite{houck2012, ma2019, yan2019, yanay2020, saxberg2022, karamlou2022, karamlou2024}.
Furthermore, recent advancements in the control of superconducting qubits allow the emulators to include a broadly adjustable magnetic vector potential~\cite{rosen2024, wang2024, chirolli2024}.

Here, we study particle localization in an array of coupled transmon qubits arranged in a chain of $2\times2$ plaquettes, forming the one-dimensional rhombic lattice.
Each plaquette is threaded by an equal and adjustable synthetic magnetic flux $\Phi$.
Adjusting $\Phi$ tunes the bandwidth, and therefore the kinetic energy of the bands.
Intriguingly, at $\Phi=\pi$ the rhombic lattice has the uncommon~\cite{neves2024} property that all of its bands become flat, meaning that flat-band behavior determines the dynamics at all energy scales.
We verify a crossover from nearly ballistic transport at $\Phi=0$ to localized behavior due to the fully-flat-band condition at $\Phi=\pi$.
We then repeat the experiment in the presence of on-site energy disorder and observe Anderson localization (disorder decreases propagation) near $\Phi=0$ and disorder-induced delocalization (disorder compromises the flat bands and increases propagation) near $\Phi=\pi$.
Surprisingly, we find evidence of single-parameter scaling collapse in the vicinity of the localization-delocalization crossover, which in some cases implies a phase transition although here we do not make such a claim~\cite{abrahams2001, dobrosavljevic2012}.
Finally, we study transport with correlated disorder profiles and with multiple interacting particles present, finding significantly altered localization patterns in both cases.

\section{Methods}

\begin{figure*}[ht!]
\subfloat{\label{fig:chip}}
\subfloat{\label{fig:lattice}}
\subfloat{\label{fig:transmon}}
\subfloat{\label{fig:bands_0}}
\subfloat{\label{fig:estates_0}}
\subfloat{\label{fig:bands_pi}}
\subfloat{\label{fig:estates_pi}}
\includegraphics{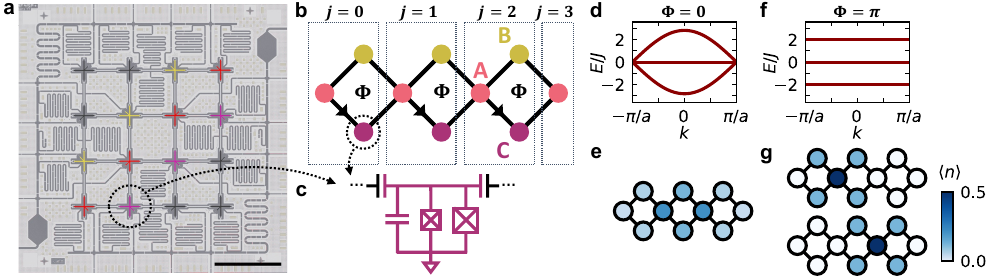}
\caption{\textbf{Realizing the rhombic lattice in a transmon qubit array.} \textbf{(a)} An optical micrograph of the 4-by-4 qubit array. The three central diagonals (false color) form three-unit-cells of the rhombic lattice, and the six remaining qubits are unused (dark grey). Scale bar, $\SI{1}{mm}$.
\textbf{(b)} A schematic of the rhombic lattice.  
Dashed boxes indicate unit cells.
Flux $\Phi$ is threaded through each 2-by-2 plaquette by inducing Peierls phases on the bonds indicated by arrows. 
\textbf{(c)} Each lattice site consists of an asymmetric tunable transmon. Nearest neighbors are capacitively coupled. 
\textbf{(d)} The continuum single-particle band structure of the rhombic lattice at $\Phi=0$. Two dispersive bands flank a flat band.
\textbf{(e)} The population distribution of the single-particle ground state of three unit cells at $\Phi=0$. The state extends over all lattice sites.
\textbf{(f)} The band structure at $\Phi=\pi$. 
\textbf{(g)} The two degenerate ground states at $\Phi=\pi$. Each state is localized to one A sublattice site and the adjacent B and C sites.
} 
\label{fig:fig1}
\end{figure*}

We emulate a tight-binding model on the rhombic lattice using 10 qubits located on the three center-most diagonals of a superconducting quantum processor having a total of 16 qubits arranged in a $4\times 4$ grid (Fig.~\ref{fig:chip}). 
The 10 qubits form three unit cells of the rhombic lattice, plus a terminating site (Fig.~\ref{fig:lattice}). 
We index the lattice sites as M$_j$, where M indicates sublattice (A, B, or C), and $j$ labels the unit cell.
Each lattice site corresponds to a flux-tunable transmon qubit (Fig.~\ref{fig:transmon}); an excitation of the qubit represents a bosonic particle occupying the site.
Nearest-neighbors are capacitively coupled, allowing excitations to tunnel between adjacent qubits.
Control lines and readout resonators are located on a separate chip and are brought into proximity of the qubits using a flip-chip configuration~\cite{rosenberg2017}.

We use a parametric-coupling scheme to generate tunable synthetic magnetic flux $\Phi$ threading each $2\times2$ plaquette.
Each qubit is energetically detuned from its neighbors.
For each nearest-neighbor pair, one qubit is modulated at a frequency matching the detuning, inducing particle-exchange interactions between the two sites.
The modulation amplitudes are chosen to yield a particle exchange rate of $J/2\pi=\SI{2}{MHz}$ for all nearest-neighbor pairs.
The flux $\Phi$ is emulated by adding a complex phase to the exchange interaction between the A and C sublattice sites in each unit cell, which is accomplished by adding an equivalent phase to the corresponding modulation tone.
Details of the parametric coupling scheme are provided in Section~S1 of the~\supp{}~\cite{sup}, and the technique is further described in Ref.~\cite{rosen2024}.
The qubits have average anharmonicity $U/2\pi=\SI{-218(6)}{MHz}\gg J/2\pi$ so that, for each site, occupation beyond a single particle is effectively disallowed.
The six unused qubits in the processor are detuned far from the active qubits to render them inactive throughout this work.

The resulting Hamiltonian is 
\begin{align}\label{eq:H}
    \frac{\hat H}{\hbar} &= J \sum_{j=0}^2 \Big( \hat{a}_{A,j}^\dag \hat{a}_{B,j} + e^{i\Phi} \hat{a}_{A,j}^\dag \hat{a}_{C,j} \nonumber\\
    &+ \hat{a}_{B,j}^\dag \hat{a}_{A,j+1} + \hat{a}_{C,j}^\dag \hat{a}_{A,j+1} \Big) + \mathrm{H.C.},
\end{align}
written in the instantaneous rotating frame of each qubit and neglecting terms that rotate at multiples of the detunings between nearest neighbors.
Here, $\hat{a}_{M,j}$ is the bosonic destruction operator for a particle occupying site $\mathrm{M}_j$, restricted to binary occupation $\hat{n}_{M,j}\equiv \hat{a}^\dag_{M,j}\hat{a}_{M,j}\in \{0, 1\}$. We neglect couplings beyond nearest neighbors, which are comparatively weak (see Section~S1 of the \supp{}).

\section{Localization in the rhombic lattice}

\begin{figure*}[ht!]
\subfloat{\label{fig:fig2_diagram}}
\subfloat{\label{fig:fig2_data}}
\subfloat{\label{fig:fig2_sim}}
\subfloat{\label{fig:fig2_moment_vs_time}}
\subfloat{\label{fig:fig2_moment}}
\subfloat{\label{fig:fig2_ipr}}
\includegraphics{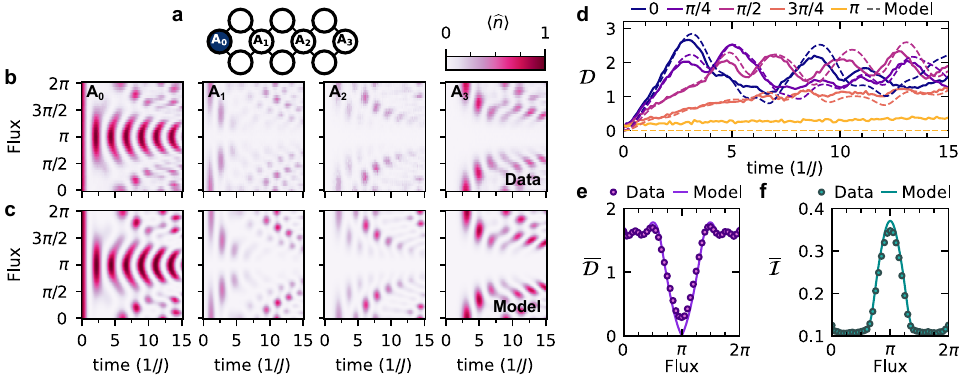}
\caption{\textbf{Flat-band localization around half-flux.} 
\textbf{(a)} A particle is initialized at the leftmost lattice site A$_0$ and is then allowed to propagate.
\textbf{(b)} The measured population of the A sublattice sites as a function of propagation time at various fluxes. 
\textbf{(c)} The simulated population of the A sublattice sites. 
\textbf{(d)} The RMS position $\mathcal{D}$ is extracted from the population within each unit cell as a function of time. Data and simulations are shown at several values of flux. 
\textbf{(e)} The RMS position of the time-averaged population, $\bar{\mathcal{D}}$, is shown as a function of flux.
\textbf{(f)} The inverse participation ratio of the time-averaged population, $\bar{\mathcal{I}}$, is shown versus flux.}
\label{fig:fig2}
\end{figure*}

The Hamiltonian (\ref{eq:H}) describes a tight-binding model for strongly-interacting bosonic particles on a rhombic lattice.
In the continuum limit, at $\Phi=0$ the rhombic lattice features two dispersive bands surrounding a zero-energy flat band (Fig.~\ref{fig:bands_0}).
As $\Phi$ increases, the bandwidth of the dispersive bands narrows while the zero-energy flat band is preserved (as shown in Section~S2 of the \supp{}).
This trend continues until, at $\Phi=\pi$, all three bands become flat, with energies $E=0,\,\pm2J$ (Fig.~\ref{fig:bands_pi}).

Although our system has finite size, the structure of its single-particle states should be related to the continuum band structure.
According to exact diagonalization of Eq.~(\ref{eq:H}), when $\Phi=0$ the ground state is extended, meaning the wavefunction has nonzero amplitude at all sites (Fig.~\ref{fig:estates_0}).
In contrast, for the flat-band condition $\Phi=\pi$, the system has two degenerate ground states, each with nonzero amplitude only at one A sublattice site and the four adjacent B and C sites (Fig.~\ref{fig:estates_pi}).
These eigenstates are called compactly localized states (CLSs).
CLSs are understood to be general features of flat bands~\cite{sutherland1986, goda2006, bodyfelt2014}, and in the rhombic lattice may be understood in terms of Aharonov-Bohm caging---destructive interference across each $2\times 2$ plaquette~\cite{vidal1998, cartwright2018, martinez2023, rosen2024}.
Since all three bands are flat at $\Phi=\pi$, all eigenstates of the Hamiltonian may be expressed as CLSs~\cite{vidal2000, mukherjee2018, gligoric2020, longhi2021}.
Note that the energy of eigenstates at the boundary of the lattice are shifted in energy from $E=\pm 2J$, but remain compactly localized.
Crucially, at $\Phi=\pi$, a single-particle excitation that is initially localized in position space overlaps only with a few nearby CLSs; therefore, the excitation cannot propagate.
When $\Phi$ is near but not exactly $\pi$, the eigenstates are extended.
Yet the mobility should remain low because the eigenstates within each band are nearly degenerate leading to slow dynamics.
Therefore, we expect a gradual crossover from ballistic transport at $\Phi=0$ to vanishing conductivity at $\Phi=\pi$.

\section{Results and discussion}

We study particle transport in the lattice at different values of $\Phi$ using single-particle quantum random walk experiments.
We initialize a particle at the left end of the lattice (site $\mathrm{A}_0$) using a microwave $\pi$ pulse and allow the particle to propagate freely through the lattice for a time $t$ (Fig.~\ref{fig:fig2_diagram}).
After the propagation time, particle exchange is suspended by removing the modulation tones, and we measure the occupation of all sites.
The expected occupation of each site is determined from the average of single-shot measurements of 4000 repeated experiments after postselection on the appropriate total excitation number, which partially mitigates state preparation, depolarization, and readout errors.
During the postselection step, we use predetermined readout confusion matrices for each qubit to further correct for second-order errors (simultaneous photon loss and erroneous excitation), which would otherwise pass the postselection criterion and bias the population distribution (details provided in Section~S1 of the \supp{}).

The population distribution as a function of $t$ and $\Phi$ is shown in Fig.~\ref{fig:fig2_data} and is compared to numerical simulation of the model Hamiltonian Eq.~(\ref{eq:H}) on a classical computer in Fig.~\ref{fig:fig2_sim}.
At $\Phi=0$ and $2\pi$, corresponding to maximally dispersive bands, the particle propagates to the right of the lattice, reaching the rightmost lattice site A$_3$, and then reflects back.
Approaching $\Phi=\pi$, corresponding to bands with decreasing bandwidth, the particle is less likely to reach the rightmost lattice sites, and instead is confined within a decreasing length around its initial position.
Finally, at $\Phi=\pi$, corresponding to the flat-band condition, the particle is confined within the first unit cell and the population elsewhere remains approximately zero.

\begin{figure*}[ht!]
\subfloat{\label{fig:fig3_diagram}}
\subfloat{\label{fig:fig3_data}}
\subfloat{\label{fig:fig3_moment}}
\subfloat{\label{fig:fig3_ipr}}
\includegraphics{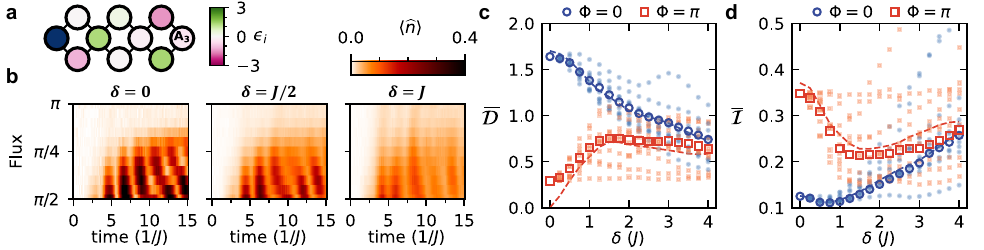}
\caption{\textbf{Anderson delocalization near half-flux.} 
\textbf{(a)} One particle is initialized at the leftmost lattice site A$_0$ in the presence of Gaussian on-site energy disorder. Profiles for on-site-energy disorder are drawn from a Gaussian distribution with unity variance; one such profile is shown by the color of each site (aside from A$_0$, the on-site-energy of which is unaltered; the profile is shown before being scaled by the disorder strength $\delta$).
\textbf{(b)} The population of the rightmost lattice site A$_3$, averaged over 10 disorder profiles, as a function of time and flux. Away from $\pi$ flux, the particle is decreasingly likely to propagate to the rightmost site with increasing $\delta$. Near $\Phi=\pi$, the particle cannot propagate to the rightmost site without disorder.
\textbf{(c,~d)}~Light-colored markers indicate $\bar{\mathcal{D}}$ and $\bar{\mathcal{I}}$ for the 10 individual disorder profiles at $\Phi=0$ and $\Phi=\pi$ as a function of disorder strength. Dark-colored markers indicate the disorder-averaged values. Dashed lines, model.}
\label{fig:fig3}
\end{figure*}

As the system is, to the extent possible, closed---not connected to a source or drain---its conductivity in the traditional sense cannot be directly probed.
Instead, to quantify transport we present the root-mean-square (RMS) position of the particle~\cite{li2022}
\begin{equation}
    \mathcal{D} = \sqrt{\sum_j j^2\langle\hat{n}_{A,j}+\hat{n}_{B,j}+\hat{n}_{C,j}\rangle}.
\end{equation}
In Fig.~\ref{fig:fig2_moment_vs_time}, we present $\mathcal{D}$ as a function of time at several values of flux.
At $\Phi=0$, $\mathcal{D}$ initially increases linearly, indicating ballistic transport, and at later times, oscillates due to reflections at the lattice boundaries.
As $\Phi$ approaches $\pi$, $\mathcal{D}$ grows less rapidly, reflecting slower propagation.
At $\Phi=\pi$, $\mathcal{D}\approx 0$ at all times, reflecting localization of the particle.

In Fig.~\ref{fig:fig2_moment}, we present $\bar{\mathcal{D}}$, defined as the RMS position of the populations of each site time-averaged over the first $\SI{1.2}{\micro s}\approx 15/J$ of propagation, to provide a singular metric quantifying the degree to which the particle is localized (smaller values indicate tighter localization).
We find that $\bar{\mathcal{D}}$ decreases as $\Phi$ approaches $\pi$, reflecting the transition towards CLSs. 
As an accompanying metric, we additionally consider the inverse participation ratio (IPR)~]~\cite{skinner1990, longhi2021}
\begin{equation}
    \mathcal{I} = \sum_{M,j} \langle \hat{n}_{M,j} \rangle^2,
\end{equation}
where larger values of $\mathcal{I}$ indicate more localization.
The IPR of the time-averaged site populations is presented in Fig.~\ref{fig:fig2_ipr}, corroborating the transition towards localization.

We next examine the dynamics of the system in the presence of on-site energy disorder.
We add uncorrelated Gaussian disorder of the form 
\begin{equation}
    H_D=\delta\sum_{j=0}^3\sum_{M} \epsilon_{M,j}\hat{a}_{M,j}^\dagger \hat{a}_{M,j},
\end{equation}
where disorder profiles are determined by sampling $\epsilon_{M,j}$ from a normal distribution with unit variance, and $\delta$ sets the disorder strength.
Without loss of generality, we set $\epsilon_{A,0}=0$ for experimental convenience.
A representative disorder profile is shown in Fig.~\ref{fig:fig3_diagram}.
We repeat experiments for 10~disorder profiles and consider the lattice site populations averaged over all profiles.
For consistency, the same 10 profiles are used for all measurements and simulations.
The profiles are shown in the \supp{}.

The disorder-averaged population of the terminal lattice site A$_3$ is shown as a function of time and flux for three disorder strengths in Fig.~\ref{fig:fig3_data}.
Away from $\Phi=\pi$, the particle reaches A$_3$ less often as disorder is increased, reflecting traditional Anderson localization.
Yet near to $\Phi=\pi$, we observe the opposite.
In the absence of disorder, the particle is forbidden from propagating through the lattice so the population of A$_3$ remains zero.
When the disorder is finite, the flat-band localization effect is partially lifted and only then do we observe nonzero population of A$_3$.
This phenomenon---decreasing localization with increasing disorder---is referred to as disorder-induced delocalization~\cite{vidal2001, goda2006, gligoric2020, longhi2021} and has been observed in a lattice of ultracold atoms~\cite{li2022}.
To further quantify delocalization, we present the time-averaged RMS position and IPR in Figs.~\ref{fig:fig3_moment} and \ref{fig:fig3_ipr}, respectively, at $\Phi=0$ and~$\pi$.
At $\Phi=0$, the particle localizes with increasing disorder strength, whereas at $\Phi=\pi$, the particle delocalizes with weak disorder.
When $\delta\gtrsim 2J$, conventional Anderson localization overcomes delocalization, and the particle again localizes with further increasing disorder.

\begin{figure*}[ht!]
\subfloat{\label{fig:fig4_crossover}}
\subfloat{\label{fig:fig4_scaling}}
\subfloat{\label{fig:fig4_diagram}}
\subfloat{\label{fig:fig4_correlated}}
\includegraphics{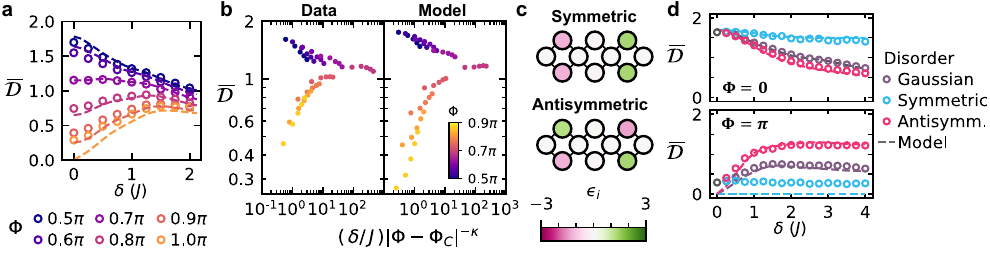}
\caption{\textbf{Localization-delocalization crossover and correlated disorder.} 
\textbf{(a)} The RMS position of the particle $\mathcal{D}$, averaged over time and 10 disorder instantiations, versus disorder strength at various fluxes. Dashed lines, model. For weak disorder $\delta<J$, we observe a crossover from Anderson localization (decreasing $\bar{\mathcal{D}}$ with increasing disorder) to Anderson delocalization (increasing $\bar{\mathcal{D}}$) as the flux approaches $\pi$. When $\Phi\approx0.7\pi$, $\bar{\mathcal{D}}$ is invariant to weak disorder.
\textbf{(b)} The same data, rescaled in terms of the single parameter $\delta |\Phi-\Phi_C|^\kappa$ with $\Phi_C=0.69\pi$ and $\kappa=1.7$. 
\textbf{(c)} Examples of symmetric and antisymmetric disorder profiles.
\textbf{(d)} The RMS position of a particle, averaged over time and 10~disorder instantiations, after initialization at the leftmost lattice site A0 in the presence of Gaussian and correlated disorder.}
\label{fig:fig4}
\end{figure*}

We note that for any $\delta>0$, the localization length is finite (see Section~S3 of the \supp{}), meaning the system scales as an insulator for all $\Phi$.
Disorder-induced delocalization increases the conductivity of the system; the system nevertheless remains insulating.
This contrasts from weak anti-localization, which enhances the conductivity of itinerant states.

To further examine the crossover between localization and delocalization, we present $\bar{\mathcal{D}}$ versus $\delta$ for several intermediate values of $\Phi$ in Fig.~\ref{fig:fig4_crossover}.
Intriguingly, for weak disorder, $\bar{\mathcal{D}}$ remains roughly constant when $\Phi=0.7\pi$.
This observation suggests a single-parameter scaling collapse analysis.
Traditionally, such analyses can reveal metal-insulator transitions in semiconductors: in the vicinity of a quantum phase transition at a critical value $n_C$ of the carrier density $n$, the resistance of the material at various $n$ and temperatures $T$ may be approximately determined based on only a single parameter of the form $T |n-n_C|^{-\kappa}$, where $n_C$ and $\kappa$ are empirically found and $\kappa$ is related to the universality class of the transition~\cite{kravchenko1995, abrahams2001, dobrosavljevic2012}.
We look for single-parameter scaling of the form $\delta |\Phi-\Phi_C|^{-\kappa}$, where $\Phi_C$ is the critical flux value.
Selecting $\Phi_C=0.69\pi$ and $\kappa=1.7$, we observe approximate collapse of the data above and below $\Phi_C$ onto single curves, as shown in Fig.~\ref{fig:fig4_scaling}.
Further numerical simulations presented in Section~S3 of the \supp{} suggest that the observation of scaling collapse is not an artifact of finite system size nor of finite time.
Yet an analysis of the system's eigenstates in the presence of disorder, presented in the same Section, yields no evidence of a critical point.
The appearance of a sudden crossover from localization to delocalization at $\Phi=0.7\pi$ therefore remains an intriguing aspect of single-particle dynamics in the rhombic lattice.

We next study dynamics in the presence of correlated disorder.
We consider symmetric disorder, where $\epsilon_{B,j} = \epsilon_{C,j}$, and antisymmetric disorder, where $\epsilon_{B,j} = -\epsilon_{C,j}$.
We use the same values of $\epsilon_{C,j}$ as used in the Gaussian disorder profiles, and we set $\epsilon_{A,j}=0$.
Symmetric and antisymmetric disorder profiles are exemplified in Fig.~\ref{fig:fig4_diagram}.

In Fig.~\ref{fig:fig4_correlated}, we compare $\bar{\mathcal{D}}$ as a function of $\delta$ for uncorrelated, symmetric, and antisymmetric disorder at $\Phi=0$ and~$\pi$.
At $\Phi=0$, we observe localization under all three types of disorder, though localization is more pronounced for uncorrelated and antisymmetric disorder than for symmetric disorder.
In Section~S4 of the \supp{}, we show analytically that there is a zero-energy extended state that is unaffected by increasing symmetric disorder, providing residual conductivity in the lattice even as the disorder strength becomes large.
This feature is reminiscent of the random dimer model, where some extended states in a one-dimensional chain are immune to dimerized disorder~\cite{phillips1990}.

At $\Phi=\pi$, we observe that delocalization is most pronounced for antisymmetric disorder.
Interestingly, the particle remains localized under symmetric disorder. 
This observation can be explained by examining the eigenstates.
In the absence of disorder, all eigenstates can be expressed as CLSs having nonzero amplitude on five lattice sites.
Under symmetric disorder, every eigenstate remains confined to the same five sites, and the system therefore remains localized~\cite{longhi2021}.
CLSs in the presence of symmetric disorder are exemplified in Section~S4 of the~\supp{}.
In contrast, at $\Phi=\pi$, delocalization is the most pronounced for antisymmetric disorder, supporting the prediction that a zero-energy extended state exists in the presence of antisymmetric disorder (whereas under uncorrelated disorder, all eigenstates are localized but not compactly localized)~\cite{longhi2021}.

\begin{figure*}[ht!]
\subfloat{\label{fig:fig5_AB}}
\subfloat{\label{fig:fig5_AC}}
\subfloat{\label{fig:fig5_BC}}
\subfloat{\label{fig:fig5_ABC}}
\includegraphics{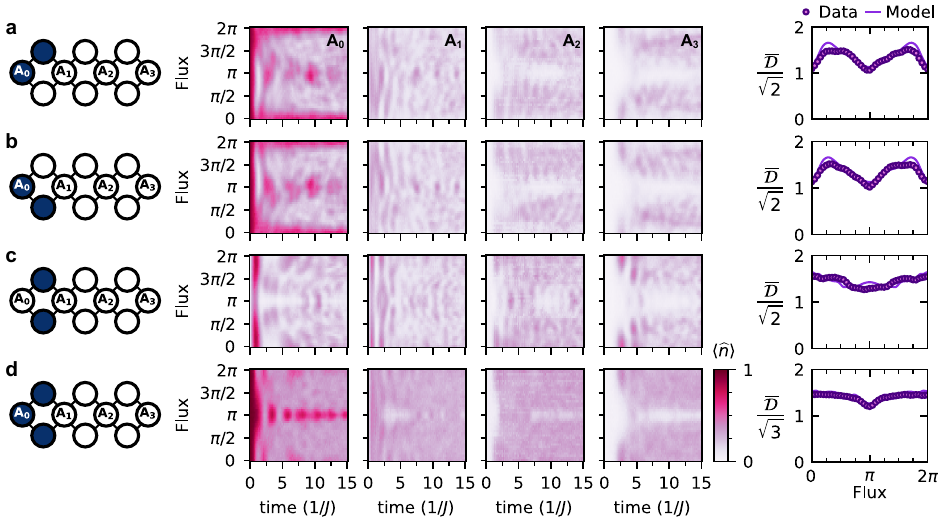}
\caption{\textbf{Dynamics of multiple interacting particles} in the absence of disorder.
Each panel shows, from left to right, a diagram of the initial state with dark blue coloring indicating the initially-filled sites, the measured population at all A sublattice sites as a function of time and flux, and the corresponding time-averaged RMS position versus flux. Positions are normalized by the total number of particles.
\textbf{(a)} Two particles are initialized the leftmost unit cell, one at the A sublattice site and one at the B sublattice site. 
\textbf{(b)} Two particles are initialized at the A and C sublattice sites.
\textbf{(c)} Two particles are initialized at the B and C sublattice sites.
\textbf{(d)} Three particles are initialized at the A, B and C sublattice sites.}
\label{fig:fig5}
\end{figure*}

Lastly, we consider the dynamics of multiple interacting particles.
So far, we have shown that at $\Phi=\pi$, destructive Aharonov-Bohm interference leads to single-particle localization, and that adding non-symmetric disorder interrupts this interference, causing delocalization.
Interactions between particles should also alter the interference, so one might expect significantly different behavior when multiple particles are present~\cite{vidal2000, creffield2010}.
In Fig.~\ref{fig:fig5}, we present the dynamics following initialization of two and three particles at various sites in the first unit cell, and we extract $\bar{\mathcal{D}}$ to quantify localization.
Because the Hubbard energy $U$ exceeds the particle exchange strength $J$ by a factor of $\approx 100$, the interactions are effectively hard-core (while particle exchange statistics remain bosonic, two particles may not occupy the same site).
For all initial states, we observe some localization near $\Phi=\pi$, in that the particles are less likely to reach the rightmost lattice site A$_3$.
Yet the particles are not fully confined within the first unit cell, as we earlier demonstrated for a single particle, implying that interactions between particles indeed disrupt the quantum interference that leads to flat single-particle bands.
Details of the dynamics depend on the initial configuration.
For example, we also observe some localization near $\Phi=0$ following initialization at A$_0$ and B$_0$, but not following initialization at B$_0$ and C$_0$.

\section{Conclusion}

In this work, we explore localization effects in an artificial flat-band lattice.
Using a superconducting qubit array, we emulate a tight-binding model on the rhombic lattice with broad parameter adjustability, and we tune between the regimes of dispersive and flat bands by controlling the synthetic magnetic flux threading each unit cell.
Transport responds to disorder differently in the two regimes.
Dispersive bands permit ballistic particle motion, and adding disorder induces conventional Anderson localization.
Flat bands, on the other hand, do not allow particles to propagate, yet we verify that adding disorder disrupts the flat band and allows particle motion.
When the disorder profile is correlated, we show that several zero-energy states are robust to the disorder, further enhancing or suppressing conduction depending on the regime.
Along with disorder, we confirm that interactions destabilize flat band states, enhancing conduction in the flat-band regime.

We observe a sudden transition in localization physics between the dispersive band and the flat band regimes.
At the transition point, propagation is approximately independent of disorder amplitude for small disorder.
Yet, at all flux values, the average spatial extent of the eigenstates monotonically decreases with increasing disorder.
We therefore understand the transition as a competition between eigenstate localization and the suppression of group velocity in flat bands.
In the vicinity of the transition, our results align with the single-parameter scaling collapse well-known from semiconductor metal-insulator transitions.
This observation---reminiscent of quantum critical scaling behavior---is remarkable given that our finite-size scaling analysis indicates no phase boundary between the regimes with disorder-induced localization and delocalization.

Our methods are straightforward to implement with larger quantum processors as they become available.
Implementations of larger rhombic lattices could be used to study predicted Luttinger liquid phases~\cite{doucot2002, cartwright2018} and topological states~\cite{kremer2020, zuo2024}.
Lastly, disorder-induced delocalization like that observed in the present work was also recently predicted in magic-angle twisted bilayer graphene (MATBG)~\cite{alcazar2024}.
Yet studying intentionally-disordered MATBG is challenging experimentally---due to the difficulty of controllable and reproducible fabrication---and computationally---as there are thousands of lattice sites per unit cell.
Ultimately, using larger-scale quantum processors to emulate such flat-band systems would enable a straightforward means to test the impact of parameter variations on the system physics without the need to fabricate each individual configuration, a task that is often practically prohibitive to achieve.

\section{Acknowledgements}

The authors are grateful to Roopayan Ghosh, Madhumita Sarkar, Gyunghun Kim, Terry P. Orlando, Lev B. Ioffe, and Leonid S. Levitov for fruitful discussions.
This material is based upon work supported by the U.S. Department of Energy, Office of Science, National Quantum Information Science Research Centers, Quantum System Accelerator. 
Additional support is acknowledged from the Defense Advanced Research Projects Agency under the Quantum Benchmarking contract, from U.S. Army Research Office Grant W911NF-23-1-0045, from the U.S. Department of Energy, Office of Science, National Quantum Information Science Research Centers, Co-design Center for Quantum Advantage (C2QA) under contract number DE-SC0012704, and from the Department of Energy and Under Secretary of Defense for Research and Engineering under Air Force Contract No. FA8702-15-D-0001.
ITR is supported by an appointment to the Intelligence Community Postdoctoral Research Fellowship Program at the Massachusetts Institute of Technology administered by Oak Ridge Institute for Science and Education (ORISE) through an interagency agreement between the U.S. Department of Energy and the Office of the Director of National Intelligence (ODNI).
SM is supported by a NASA Space Technology Research Fellowship.
CNB acknowledges support from the National Science Foundation under award DMR-2141064.
Any opinions, findings, conclusions, or recommendations expressed in this material are those of the author(s) and do not necessarily reflect the views of the Department of Energy, the Department of Defense, or the Under Secretary of Defense for Research and Engineering.

\section{Author Contributions}

SM and MS designed the device.
ITR performed the experiments and analyzed the data.
SM, CNB, and DAR developed experimental infrastructure.
RD, DKK, and BMN fabricated the device with coordination from KS, MES, and JLY.
JAG and WDO provided technical oversight and support.
ITR wrote the manuscript with contributions from all authors.

\section{Competing Interests}

The Authors declare no competing interests.

\section{Data Availability}
The data that support the findings of this study are available from the corresponding author on reasonable request and with the cognizance of our U.S. Government sponsors who financed the work.

\nocite{barrett2023, karamlou2024, bukov2015, alaeian2019, zhao2022, wang2024, cartwright2018, skinner1990, sarkar2023}
\bibliography{refs}



\section*{Supplementary material}

## S1. EXPERIMENTAL DETAILS

### A. Device parameters

Our experiment is conducted in a dilution refrigerator at a base temperature of approximately 22 mK. The device is an array of 16 capacitively-coupled flux-tunable transmon qubits, 10 of which are used in the present experiment. The 10 qubits that are used are flux-biased to excitation frequencies between 4.5 GHz and 4.8 GHz; the six unused qubits are biased to 4.15 GHz to prevent interactions with the active qubits. The qubits are arranged in a 4-by-4 square lattice. Nearest-neighboring qubits are capacitively coupled to realize bare exchange couplings with average strength $J_0/2\pi = 5.9$ MHz and standard deviation 0.4 MHz. The average bare exchange coupling strength between next-nearest neighbors is 0.43 MHz with standard deviation 0.23 MHz, however the parametric coupling scheme largely mitigates the impact of beyond-nearest-neighbor couplings.

Qubit control lines and readout resonators are located on a separate chip and are brought in proximity to the qubits via a flip-chip process. Each qubit is capacitively coupled to a co-planar waveguide resonator for dispersive readout. Control pulses, including DC flux biases, parametric flux modulation tones, and excitation pulses resonant with qubit excitation frequencies, are sent through individual control lines coupled to each qubit. The circuitry of the qubit tier and the control/readout tier are formed by evaporating superconducting aluminum on a silicon substrate. The chip used in the present study is discussed in more depth in Refs. [1, 2]. A diagram of the control and readout electronics used to operate the chip is shown in Fig. S1, and several experimental settings and performance metrics are described in Table SI.

### B. Parametric coupling scheme

In this work, to emulate a rhombic lattice with an adjustable flux threading each plaquette, we use a parametric coupling scheme to induce the appropriate interactions between neighboring qubits. Here, we summarize the important aspects of the coupling scheme; a full description may be found in Ref. [3].

Considering two adjacent qubits $i$ and $j$ alone, in a typical analog simulation experiment, the qubits would be tuned to the same DC frequency. Doing so would realize exchange coupling of the form $J_0(\hat{a}_i^\dagger \hat{a}_j + \hat{a}_i \hat{a}_j^\dagger)$, where the bare coupling strength $J_0$ is determined by their mutual capacitance. In the parametric coupling scheme, we instead detune the two qubits and parametrically modulate one qubit at a frequency matching the detuning. The resulting lab frame Hamiltonian is

$$\hat{H}_L^{ij}/\hbar = \left(\omega + \Omega\sin(\gamma t + \phi)\right)\hat{n}_i + (\omega - \gamma)\hat{n}_j + J_0(\hat{a}_i^\dagger \hat{a}_j + \hat{a}_i \hat{a}_j^\dagger), \tag{S1}$$

where $\Omega$ is the modulation amplitude, $\phi$ is the modulation phase, and $\gamma$ is the modulation frequency and the detuning. In the instantaneous rotating frame of both qubits, and neglecting rotating terms, this Hamiltonian is

$$\hat{H}_R^{ij}/\hbar = J_0 \mathcal{J}_1\left(\frac{\Omega}{\gamma}\right)\left(e^{-i\phi}\hat{a}_i^\dagger \hat{a}_j + e^{i\phi}\hat{a}_i \hat{a}_j^\dagger\right), \tag{S2}$$

where $\mathcal{J}_n$ is the $n$th-order Bessel function of the first kind. The rotating-frame Hamiltonian reveals that the parametric modulations induce exchange coupling with strength $J = J_0 \mathcal{J}_1\left(\frac{\Omega}{\gamma}\right)$ and Peierls phase $\phi$.

We extend this scheme throughout the lattice by modulating adjacent qubits at inequivalent frequencies and setting the DC frequencies of the qubits to provide corresponding detunings. The layout of qubit frequencies, detunings, and modulation frequencies, is shown in Fig. S2. Modulating all qubits reduces the effective coupling strength: if qubit $i$ is modulated with amplitude $\Omega_i$ and frequency $\gamma_i$, and qubit $j$ is detuned by $\gamma_i$ and modulated with amplitude $\Omega_j$ and frequency $\gamma_j$, then the effective coupling strength is $J = J_0 \mathcal{J}_1\left(\frac{\Omega_i}{\gamma_i}\right)\mathcal{J}_0\left(\frac{\Omega_j}{\gamma_j}\right)$. We select modulation amplitudes to provide $J/2\pi = 2.0$ MHz, taking this effect into account.

Importantly, selecting nonzero Peierls phases emulates perpendicular magnetic flux threading the lattice. In particular, for a plaquette $P$, the Peierls phases $\phi_i$ of the bonds surrounding the plaquette are equivalent to a dimensionless flux $\Phi = \sum_{\partial P} \phi_i$ where $\partial P$ is the oriented path around $P$. Note the value of $\Phi$ is defined modulo $2\pi$. The dynamics of qubit excitations in our experiment therefore emulate the dynamics of charged particles with flux quanta $\Phi_0$ moving in a perpendicular magnetic field $\mathbf{B}$ with strength set by $\Phi_P = \frac{1}{\Phi_0}\iint_P \nabla \times \mathbf{B} \cdot d\mathbf{S}$.

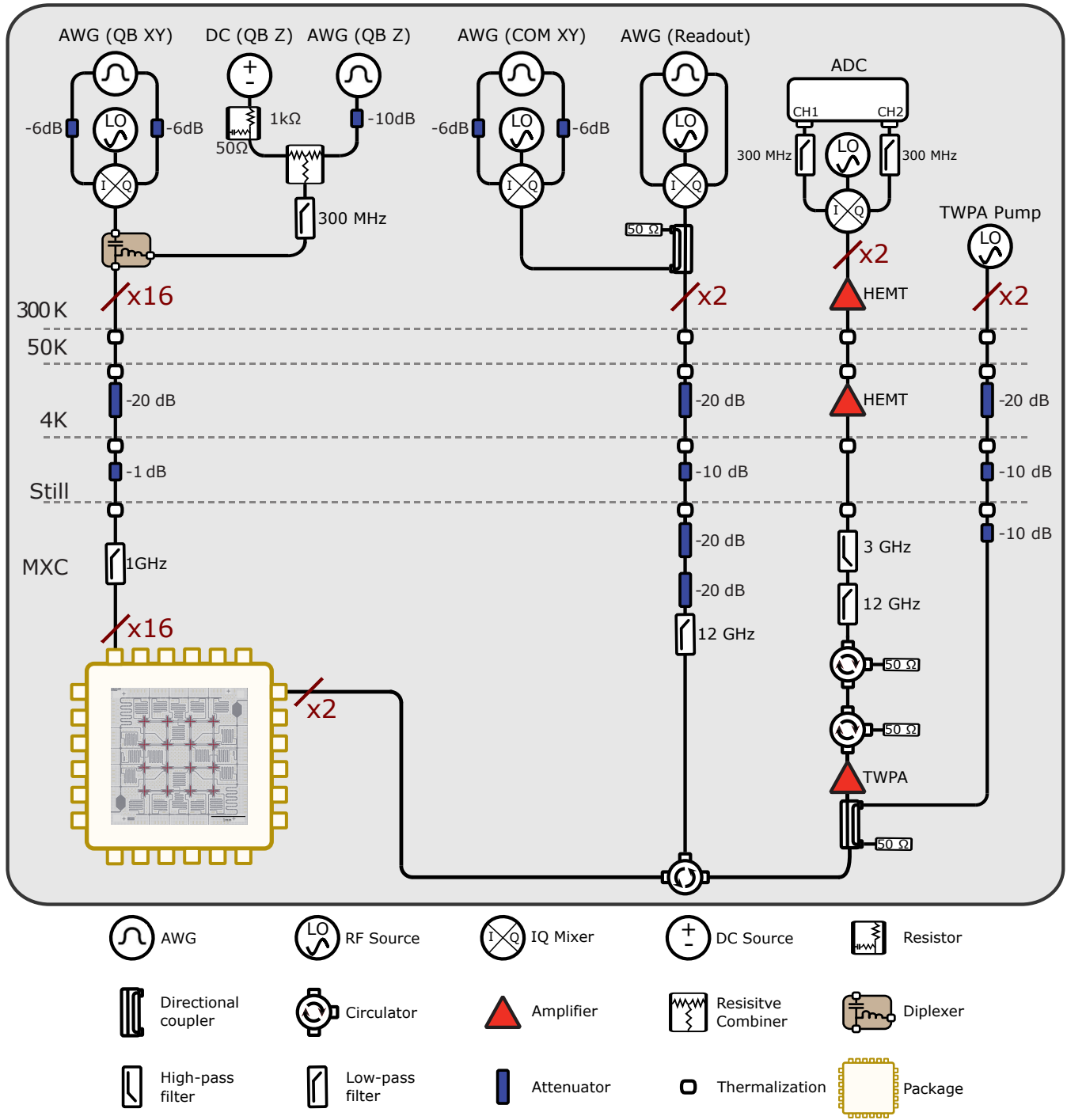

FIG. S1. **Diagram of qubit control and readout electronics** (adapted from Ref. [2].

## C. Disorder profiles

The 10 disorder profiles used in this work are shown in Fig. S3. Disorder profiles were obtained by drawing random numbers from a Gaussian distribution with unit variance.

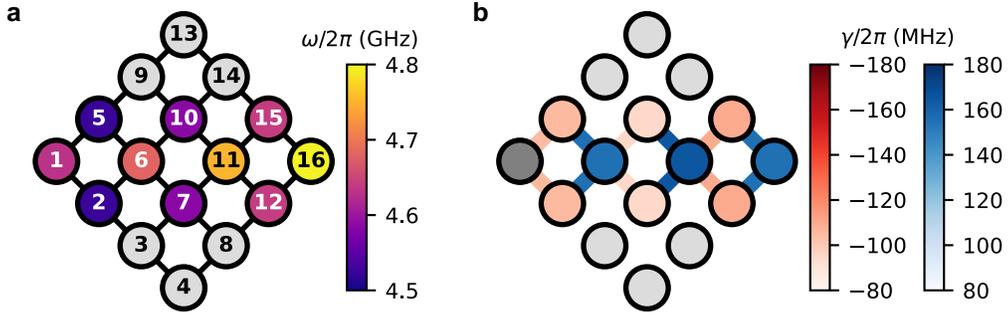

FIG. S2. **Layout for the parametric coupling scheme. (a)** A schematic of the $4 \times 4$ array, with each circle representing a qubit. Their colors describe the DC frequency setpoints of the 10 qubits used in this work, with orientation matching diagrams in the main text. The six unused qubits are shown in light grey. **(b)** The color of each circle represents the frequency at which the correponding qubit is modulated, and the color of each nearest-neighbor bond represents the detuning between the two qubits it connects. Moving from left to right, red (blue) shades indicate negative (positive) detunings as indicated by the left (right) colorbar. Qubit 1 (indicated in dark grey) is not modulated.

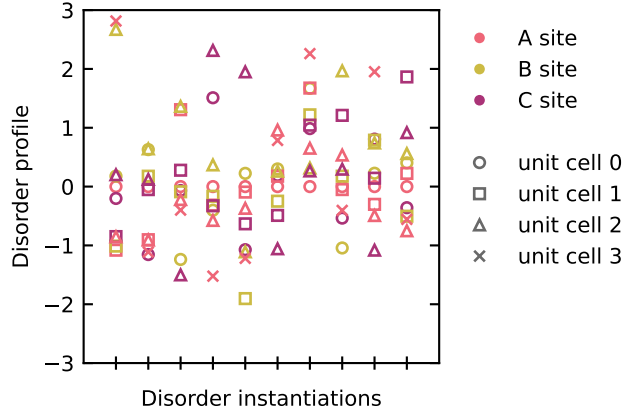

FIG. S3. **The disorder profiles used in the present work.** For each of the 10 instantiations of disorder, the value of the on-site energy disorder before being scaled by the disorder strength $\delta$ is shown for each lattice site. Lattice sites are denoted by marker colors (indicating sublattice index) and shapes (indicating unit cell index).

### D. Readout corrections

The experimental data presented in this work are based on simultaneous single-shot population measurements of the 10 qubits forming the rhombic lattice. The population on each site is determined by averaging data from 4000 single-shot measurements of identically-prepared systems. We use a two-step process to partially mitigate erroneous readout shots due to thermal population, qubit relaxation, and readout infidelity. First, we post-select each single-shot measurement on total population, discarding any measurements where the total population does match the number of particles initialized in the lattice (1, 2, or 3 particles for the data presented in this work). Measurements where one readout error occurred do not pass post-selection. However, occasionally a measurement with multiple simultaneous errors could pass post-selection; notably, an experimental repetition where the system gains an additional particle due to a thermal excitation and loses a particle due to qubit relaxation would result in an erroneous measurement that passes post-selection.

To mitigate such second-order errors, we experimentally determine the single-qubit confusion matrices for each qubit whose elements $P_{ab}^i$ are the probability of measuring qubit $i$ in the ground ($b = g$) and excited ($b = e$) states after preparing the qubit in the ground ($a = g$) and excited ($a = e$) states, such that $P_{gg}^i + P_{ge}^i = P_{ee}^i + P_{eg}^i = 1$. Assuming a uniform prior distribution for the ideal measurement outcome, the probability of a measurement where one particle is lost is

$$P_L = m\rho_L - \binom{m}{2}\rho_L^2, \tag{S3}$$

where

$$\rho_L = \frac{1}{N} \sum_{i=1}^{N} P_{eg}^i \tag{S4}$$

is the probability of losing each particle, $N = 10$ is the total number of qubits in the system, and $m$ is the correct number of particles in the system. The probability of a measurement where one particle is gained is

$$P_G = \left( \sum_{i=1}^{N} \frac{P_{ge}^i}{P_{gg}^i} \right) \left( \prod_{i=1}^{N} P_{gg}^i \right). \tag{S5}$$

The probability of a measurement where one particle is erroneously gained and a particle is erroneously lost is then

$$P_{GL} = P_G P_L = \left( \prod_{i=1}^{N} P_{gg}^i \right) \left( \sum_{i=1}^{N} P_{eg}^i \right) \left( \sum_{i=1}^{N} \frac{P_{ge}^i}{P_{gg}^i} \right), \tag{S6}$$

where the rightmost expression applies for $m = 1$. For our system, we find $P_{GL} = 0.0203$ for $m = 1$. To mitigate these second-order errors, from the population of each qubit $j$ after post-selection we subtract the probability of losing a photon and gaining a photon at qubit $j$

$$P_{GL}^j = P_L P_{ge}^j \left( \prod_{i \neq j}^{N} P_{gg}^i \right), \tag{S7}$$

and then renormalize the population distribution.

| Parameters | QB1 | QB5 | QB2 | QB6 | QB10 | QB7 | QB11 | QB15 | QB12 | QB16 |
|---|---|---|---|---|---|---|---|---|---|---|
| Lattice site | $A_0$ | $B_0$ | $C_0$ | $A_1$ | $B_1$ | $C_1$ | $A_2$ | $B_2$ | $C_2$ | $A_3$ |
| Feedline | A | A | A | A | A | B | B | B | B | B |
| $\omega_{\mathrm{res}}/2\pi$ (GHz) | 6.206 | 6.365 | 6.358 | 6.339 | 6.429 | 6.428 | 6.338 | 6.361 | 6.357 | 6.197 |
| $\omega^{\mathrm{max}}/2\pi$ (GHz) | 4.859 | 4.691 | 4.873 | 4.825 | 4.967 | 4.695 | 4.894 | 4.771 | 4.838 | 4.947 |
| $\omega/2\pi$ (GHz) | 4.630 | 4.525 | 4.525 | 4.680 | 4.585 | 4.585 | 4.750 | 4.640 | 4.640 | 4.795 |
| $\gamma/2\pi$ (MHz) | N.A. | 105 | 105 | 155 | 95 | 95 | 165 | 110 | 110 | 155 |
| $T_1$ (μs) | 25.7 | 8.0 | 17.4 | 15.9 | 12.0 | 11.4 | 18.3 | 17.1 | 17.4 | 13.3 |
| $T_{2R}$ (μs) | 3.4 | 2.9 | 2.7 | 3.9 | 1.2 | 1.9 | 1.3 | 2.0 | 2.3 | 2.4 |
| $T_\phi$ (μs) | 9.8 | 10.8 | 8.5 | 12.2 | 3.7 | 5.7 | 2.9 | 5.4 | 6.3 | 6.7 |
| $\mathcal{F}_{\mathrm{gg}}$ | 0.967 | 0.960 | 0.958 | 0.962 | 0.962 | 0.953 | 0.967 | 0.963 | 0.962 | 0.965 |
| $\mathcal{F}_{\mathrm{ee}}$ | 0.916 | 0.932 | 0.928 | 0.940 | 0.927 | 0.876 | 0.937 | 0.930 | 0.888 | 0.939 |

TABLE SI. **Summary of measurement parameters and device performance**. Performance metrics are measured at the frequency setpoints used in the present experiment. The following are shown: the lattice site the qubit represents, to which of the two readout resonator feedlines each qubit is linked, the readout resonator frequencies $\omega_{\mathrm{res}}$, the maximum transmon transition frequencies $\omega^{\mathrm{max}}$ at the upper flux-insensitive point, the transmon transition frequency setpoints for the experiment $\omega$, the modulation frequency $\gamma$ applied to each qubit for parametric coupling, the measured qubit decay times $T_1$, Ramsey coherence times $T_{2R}$, dephasing times $T_\phi$ extracted from Hahn echo sequences, and the measurement fidelities $\mathcal{F}_{gg}$ ($\mathcal{F}_{ee}$) of measuring the qubit the ground (excited) state after preparing it in the ground (excited) state. Readout fidelities are primarily limited by state preparation error due to thermal qubit population; $\mathcal{F}_{ee}$ is additionally limited by decay during readout.

## S2. RHOMBIC LATTICE: BAND STRUCTURE

Fig. S4 presents the single-particle band structure of the continuum rhombic lattice, calculated analytically [4]. Importantly, the bandwidth of the non-zero-energy bands decreases continuously as the flux is tuned from 0 to $\pi$, where all bands become flat.

The theoretical single-particle eigenspectrum of an ideal three-unit-cell rhombic lattice, obtained by exact diagonalization, is shown as a function of flux in Fig. S5a. The zero-energy flat band is clearly visible at all flux values, and flat bands at $\Phi = \pi$ appear as degenerate states at $E = \pm 2J$. The four other states at $\Phi = \pi$ are compactly localized states at the left and right edges of the lattice. These states are similar to the states at $E = \pm 2J$, but their energy is altered due to boundary effects. To clarify, in Fig. S5b we provide the single-particle eigenspectrum of a twelve-unit-cell lattice, where the bulk and boundary states become clearly distinct at $\Phi = \pi$. The spectra of two- and three-particle states in a three-unit-cell lattice are shown in Fig. S5c and S5d, respectively. While there are highly-degenerate states at zero energy, there are also states at extensively many independent energies, reflecting that at no value of $\Phi$ are all eigenstates compactly localized.

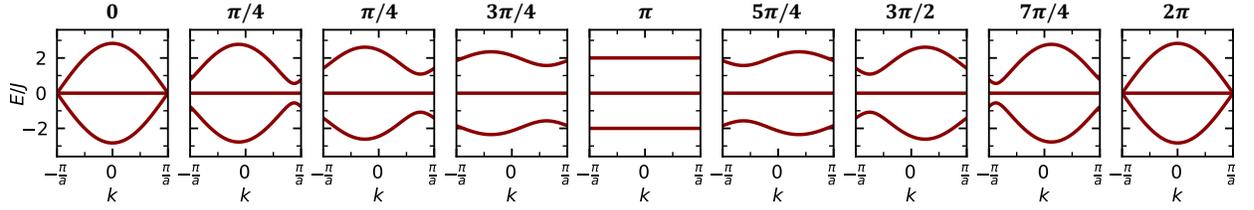

FIG. S4. **Continuum single-particle band structure of the rhombic lattice.** The three bands are shown at various values of the flux $\Phi$ threading each plaquette, as indicated at top. Momenta are shown in terms of the lattice constant $a$.

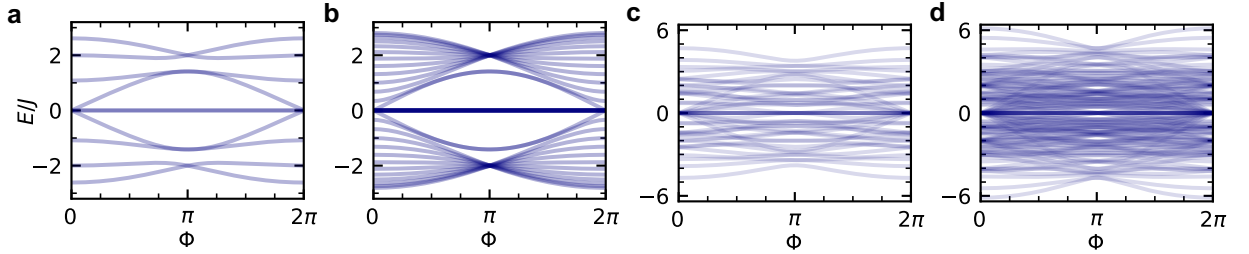

FIG. S5. **Eigenspectra of the rhombic lattice at finite size**, shown as a function of flux $\Phi$. **(a)** The single-particle eigenenergies of a three-unit-cell length rhombic lattice, as studied in the present work. **(b)** The single-particle eigenenergies of a 12-unit-cell length rhombic lattice, shown to emphasize the zero-energy flat band and the transition between dispersive bands at $\Phi = 0$ to flat bands at $\Phi = \pi$. **(c)** The two-particle eigenenergies of a three-unit-cell length rhombic lattice with hard-core interactions. **(d)** The three-particle eigenenergies of a three-unit-cell length rhombic lattice with hard-core interactions.

## S3. SCALING ANALYSIS

Here, using numerically computed exact diagonalization, we study the system-size scaling behavior of the single-particle eigenstates of the rhombic lattice. In Fig. S6, we present the mean inverse participation ratio (IPR) $\mathcal{I}$ of all single-particle eigenstates for three system sizes. For all system sizes, and at all values of $\Phi$, $\mathcal{I}$ monotonically increases with increasing disorder, suggesting that the single-particle eigenspectrum scales everywhere as an Anderson insulator. Note that values are not shown at $\delta = 0$ because the mean IPR of the eigenstates is not well-defined when the eigenstates are highly degenerate.

We further support this conclusion in Fig. S7 where, following Refs. [5, 6], we compute the fractal dimension of the eigenstates as a function of flux and disorder strength. We compute the average inverse participation ratio of all eigenstates of systems of size $2^m$ unit cells for $m = 4$ through $m = 10$. We then fit these data to a power law; examples are shown in Fig. S7a. The fractal dimension $D_2$ is defined as $-1$ times the exponent of the power law, and is presented in Fig. S7b. Ergodic states appear as $D_2 = 1$, while insulating states appear as $D_2 = 0$. Intermediate values near the ergodic phase are an artifact of the finite size of the simulated systems; the transition becomes sharper as larger system sizes are considered. These results imply that the single-particle band structure of the rhombic lattice scales as an insulator for any finite disorder, regardless of $\Phi$.

Figs. S8 and S9 provide evidence that the observed effect of delocalization is not an artifact of finite size nor of finite time. In Fig. S8a, we present the time-averaged second position moment $\bar{\mathcal{D}}$ over duration $15/J$ of time evolution for a three-unit-cell lattice. Delocalization appears as a region around $\Phi = \pi$ of increasing $\bar{\mathcal{D}}$ with increasing disorder strength. In Fig. S8b, we present the steady-state root-mean-squared (RMS) position $D_S$, defined as

$$D_S = \sqrt{\sum_i \langle \varepsilon_i | \hat{\mathcal{V}} | \varepsilon_i \rangle \left| \langle \varepsilon_i | \psi_0 \rangle \right|^2},$$ (S8)

where $|\psi_0\rangle$ is the initial state of particle, $|\varepsilon_i\rangle$ are the eigenstates, and the mean squared position operator is defined as

$$\hat{\mathcal{V}} = \sum_j j^2 (\hat{n}_{A,j} + \hat{n}_{B,j} + \hat{n}_{C,j}).$$ (S9)

Delocalization appears in a similar region, suggesting that delocalization is not a time-transient effect. In Fig. S9, we present analogous results for a 12-unit-cell lattice. Delocalization appears in similar regions, suggesting that delocalization is not an effect of finite size.

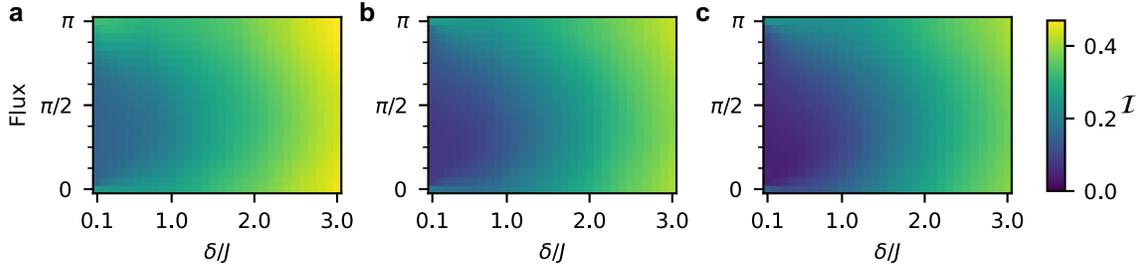

FIG. S6. **Average inverse participation ratio of single-particle eigenstates** as a function of flux and disorder strength $\delta$. Values are averaged over 10 Gaussian-distributed disorder instantiations. Results are shown for systems of length **(a)** 3, **(b)** 12, and **(c)** 48 unit cells.

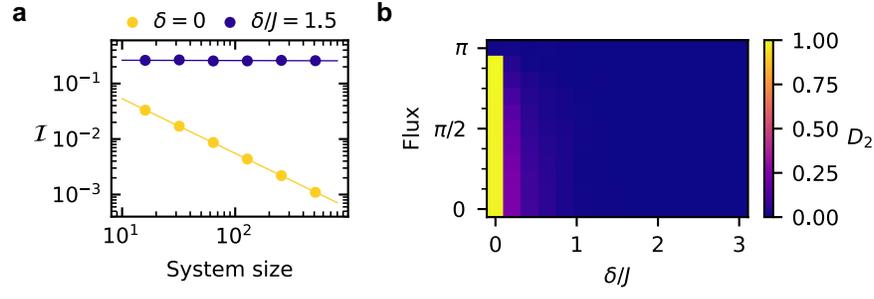

FIG. S7. **Fractal dimension of the rhombic lattice** as a function of flux $\Phi$ and disorder strength $\delta$. Values are averaged over 10 Gaussian-distributed disorder instantiations. **(a)** The average inverse participation ratio of all single-particle eigenstates, shown as a function of system size. Data is presented for $\delta = 0$ and $\delta = 1.5J$ with $\Phi = 0$, and power-law fits are shown by solid lines. **(b)** The fractal dimension $D_2$.

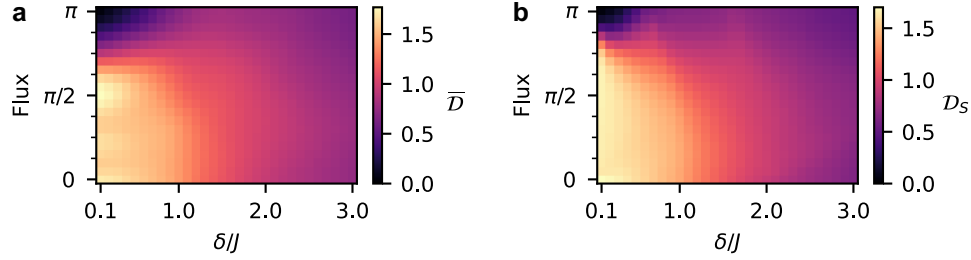

FIG. S8. **RMS position after initializing a particle at the leftmost lattice site** in a three-unit-cell size system. Values are averaged over 10 Gaussian-distributed disorder instantiations. **(a)** The time-averaged RMS position. **(b)** The steady-state RMS position.

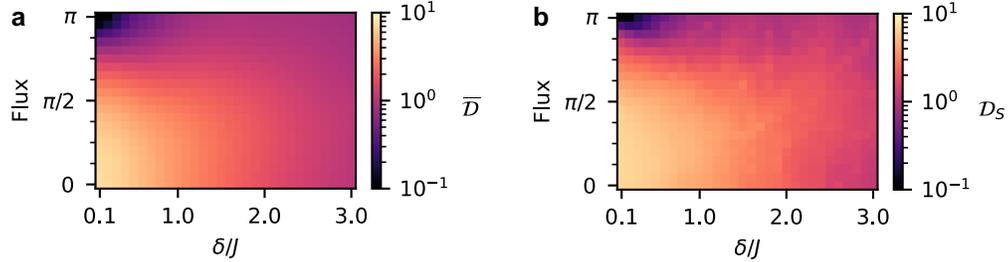

FIG. S9. **RMS position after initializing a particle at the leftmost lattice site** in a 12-unit-cell size system. Values are averaged over 10 Gaussian-distributed disorder instantiations. **(a)** The time-averaged RMS position. **(b)** The steady-state RMS position.

## S4. CORRELATED DISORDER

The Hamiltonian of the rhombic lattice (Eq. (1) of the main text) yields single-particle dynamics given by the differential equations

$$
\begin{aligned}
i\dot{A}_n &= J(B_{n-1} + e^{i\Phi}C_{n-1} + B_n + C_n) + \epsilon_{A,n}A_n \\
i\dot{B}_n &= J(A_n + A_{n+1}) + \epsilon_{B,n}B_n \\
i\dot{C}_n &= J(A_n + e^{-i\Phi}A_{n+1}) + \epsilon_{C,n}C_n,
\end{aligned}
\tag{S10}
$$

where $A_n$, $B_n$, and $C_n$ are the probability amplitudes for the particle occupying the A, B, and C sublattice sites of the $n$th unit cell, respectively, and $\epsilon_{M,n}$ are the corresponding on-site disorder energies. In this section, we describe single-particle eigenstates of the system, which appear as solutions of Eq. (S10) in the form $M_n \propto e^{i\varepsilon t}$ where $\varepsilon$ is the energy.

The flat bands correspond to eigenstates with compactly localized support, i.e., eigenstates with nonzero amplitudes in the $j-1$ and $j$ unit cells and zero amplitude elsewhere for each particular site index $n = j$. In the absence of disorder (all $\epsilon_{M,n} = 0$), it is straightforward to show that for all fluxes $\Phi$ the flat band at energy $\varepsilon = 0$ corresponds to the eigenstates with coefficients $A_j = 0$, $B_{j-1} = \frac{1}{2}$, $C_{j-1} = -\frac{1}{2}e^{i\Phi}$, $B_j = \frac{1}{2}e^{i\Phi}$, $C_j = -\frac{1}{2}$, and zero amplitude elsewhere for each site index $j$.

In addition, at $\Phi = \pi$, there are solutions with coefficients $A_j = \pm\frac{1}{\sqrt{2}}e^{\pm 2iJt}$, $B_{j-1} = -\frac{1}{2\sqrt{2}}e^{\pm 2iJt}$, $C_{j-1} = \frac{1}{2\sqrt{2}}e^{\pm 2iJt}$, $B_j = -\frac{1}{2\sqrt{2}}e^{\pm 2iJt}$, $C_j = -\frac{1}{2\sqrt{2}}e^{\pm 2iJt}$, corresponding to the flat bands at energy $\varepsilon = \pm 2J$.

### A. Symmetric disorder at $\Phi = 0$

In the main text, we show that disorder-induced localization at $\Phi = 0$ is less pronounced for symmetric disorder profiles. This is a consequence of a zero-energy extended state that is resilient to symmetric disorder, and therefore leads to residual conductivity even in the presence of strong symmetric disorder. In particular, it is straightforward to see that at $\Phi = 0$, there is a $\varepsilon = 0$ solution to Eq. S10 with non-normalized coefficients $A_n = (-1)^n$, $B_n = C_n = 0$. Because symmetric disorder has $\epsilon_{A,n} = 0$, this eigenstate is unaffected by arbitrarily strong symmetric disorder.

### B. Antisymmetric disorder at $\Phi = 0$

The zero-energy eigenstate $A_n = (-1)^n$, $B_n = C_n = 0$ is also unaffected by arbitrarily strong antisymmetric disorder, as antisymmetric disorder also has $\epsilon_{A,n} = 0$. One may therefore wonder why do we not observe residual conductivity in the case of antisymmetric disorder. The reason is that, under antisymmetric disorder, this state is part of flat band of many zero-energy states (these other states are pinned to zero energy under antisymmetric disorder, but not under symmetric or uncorrelated disorder). As such, all states in this zero-energy band may be expressed as CLS and therefore the band does not contribute any conductivity. These flat-band states can be found by expressing Eq. (S10) at $\Phi = 0$, zero energy, and with antisymmetric disorder:

$$
\begin{aligned}
0 &= J(B_{n-1} + C_{n-1} + B_n + C_n) \\
0 &= J(A_n + A_{n+1}) + \epsilon_n B_n \\
0 &= J(A_n + A_{n+1}) - \epsilon_n C_n,
\end{aligned}
\tag{S11}
$$

the solutions of which can be written as CLS with non-normalized coefficients $A_j = (-\epsilon_j/J)B_j = (\epsilon_j/J)C_j$ for each site index $n = j$ and zero amplitude elsewhere. Note that in the limit of zero disorder $\epsilon_n \to 0$, the A site amplitudes vanish and these CLS become equivalent to the zero-disorder, zero-energy eigenstates described earlier.

### C. Symmetric disorder at $\Phi = \pi$

In the main text, we show that there is no disorder-induced delocalization at $\Phi = \pi$ for symmetric disorder profiles. This occurs because, at $\Phi = \pi$, all eigenstates remain compactly localized under symmetric disorder, even though their wavefunctions and energies are affected. This is most easily shown numerically: in Fig. S10, we present the probability densities of the eigenstates in the unit cells surrounding a symmetric defect, obtained via exact diagonalization. The eigenstates centered at site $A_j$ have zero amplitude at sites $A_{j\pm1}$, confirming that the eigenstates remain compactly localized in the presence of the defect.

## D. Antisymmetric disorder at $\Phi = \pi$

In the main text, we show that there is rapid disorder-induced delocalization at $\Phi = \pi$ for antisymmetric disorder profiles. This effect is the result of zero-energy extended states that exist under antisymmetric disorder provided that the disorder amplitude is nonzero. These zero-energy states can be found by expressing Eq. (S10) at $\Phi = \pi$, zero energy, and with antisymmetric disorder:

$$
\begin{aligned}
0 &= J(B_{n-1} - C_{n-1} + B_n + C_n) \\
0 &= J(A_n + A_{n+1}) + \epsilon_n B_n \\
0 &= J(A_n - A_{n+1}) - \epsilon_n C_n.
\end{aligned}
\tag{S12}
$$

These equations have two linearly independent solutions, which may be defined through the recursive relationship

$$
0 = \frac{1}{\epsilon_{n-1}} A_{n-1} + \frac{1}{\epsilon_n} A_{n+1},
\tag{S13}
$$

and by setting an initial amplitude for sites of even and odd index (i.e. values of $A_{2j}$ and $A_{2j+1}$ for some $j$) [7]. The corresponding B and C sublattice site amplitudes are $B_n = -(J/\epsilon_n)(A_n + A_{n+1})$ and $C_n = (J/\epsilon_n)(A_n - A_{n+1})$. We note that these two states only exist at strictly zero-energy in the limit of an infinitely long lattice. Nevertheless, finite-sized systems do feature states close to zero energy with nonzero amplitude across many lattice sites, therefore particle propagation is supported in lattices with antisymmetric disorder.

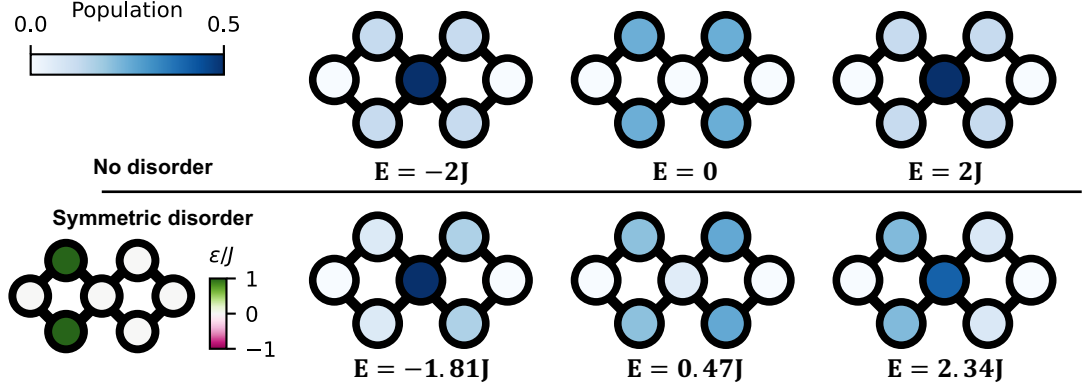

FIG. S10. **Eigenstates remain compactly localized in the presence of a symmetric disorder** at $\Phi = \pi$. Two unit cells adjacent to a symmetric defect are shown, representing the minimum system size for this demonstration. At top, the CLSs centered at the middle A sublattice site belonging to the three bands are shown at zero disorder. A symmetric defect with strength $\epsilon_j = J$ is shown at bottom left. At bottom, the corresponding eigenstates in the presence of the defect are shown. The eigenstates have exactly zero amplitude at both terminal A sublattice sites, verifying that they remain compactly localized. Eigenstate energies are indicated below each plot of their probability density.

## S5. METRICS OF PARTICLE PROPAGATION

In this section, we compare four metrics for quantifying localization through particle dynamics. In the main text, we presented the RMS position $\overline{\mathcal{D}}$ and the inverse participation ratio $\overline{\mathcal{I}}$ of the time-averaged populations of each site. In addition, we here show the effective localization length $\xi$ computed by fitting an exponential of the form $\exp(-j/\xi)$ to the time-averaged populations of each A sublattice site $A_j$. Only values $\xi \leq 3$ are shown; localization lengths exceeding the system size cannot be meaningfully determined. The effective localization length serves as a proxy for the mean decay length of single-particle eigenstates, which decay exponentially for Anderson insulators. Lastly, we define the effective conductivity $G$ as

$$G = \frac{T}{T + R}, \tag{S14}$$

where $T$ is the time-averaged population of site $A_3$, representing transmission probability, and $R$ is time-averaged population of site $A_0$, representing reflection probability.

In Fig. S11, we present the data described in Fig. 2 of the main text—the dynamics of a particle traveling in the lattice as a function of flux, without intentionally added disorder—in terms of these four metrics. In Fig. S12, we present the data described in Fig. 4a of the main text—the dynamics of a particle traveling in the lattice as a function of flux and disorder strength, averaged across 10 disorder instantiations—in terms of these four metrics. In Fig. S13, we repeat the single-parameter scaling analysis for each of the metrics. Note that, for the different metrics, we observed single-parameter scaling collapse at slightly different critical flux values $\Phi_C$; values are described in the figure caption.

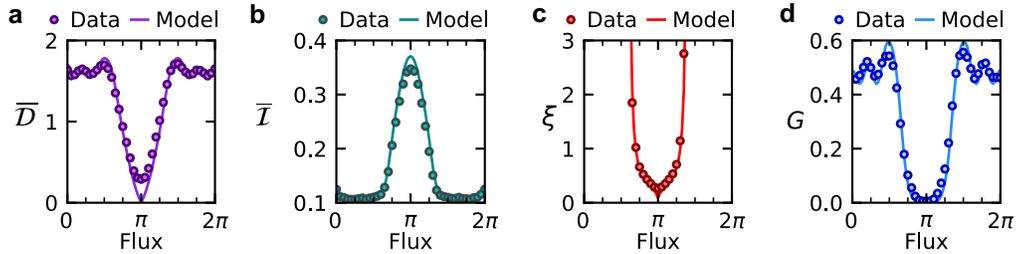

FIG. S11. **Comparing metrics of particle propagation**. Values are based on the data described in Fig. 2 of the main text, where, for various fluxes and without intentional disorder, a particle is initialized at the leftmost site (site $A_0$) and then allowed to propagate through the lattice. **(a)** The RMS position of the time-averaged population. **(b)** The inverse participation ratio of the time-averaged population. **(c)** The localization length, determined by an exponential fit of the time-averaged A sublattice site populations versus distance. **(d)** The conductivity, determined based on the time-averaged probability of the particle occupying the rightmost and leftmost sites.

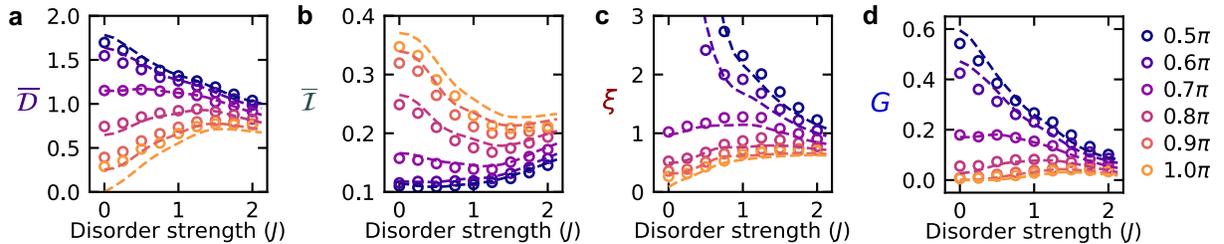

FIG. S12. **Visualizing the localization-delocalization crossover using several metrics of particle propagation**. Values are based on the data described in Figs. 4a of the main text, where the site populations are determined for 10 disorder instantiations at various fluxes and disorder strengths. **(a)** The RMS position of the time- and disorder-averaged population. **(b)** The inverse participation ratio of the time- and disorder-averaged population. **(c)** The localization length, determined by an exponential fit of the time- and disorder-averaged A sublattice site populations versus distance. **(d)** The conductivity, determined based on the time- and disorder-averaged probability of the particle occupying the rightmost and leftmost sites.

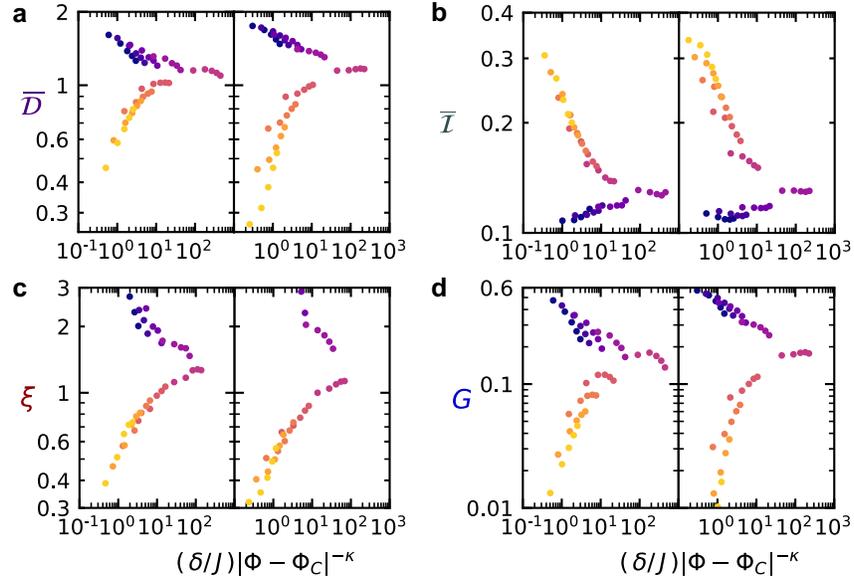

FIG. S13. **Single-parameter scaling for several metrics of particle propagation**. The data presented in Fig. S12, rescaled as a function of the single parameter $\delta|\phi - \phi_C|^\kappa$ with $\kappa = 1.7$. **(a)** The RMS position of the time- and disorder-averaged population, with $\phi_C = 0.69\pi$. **(b)** The inverse participation ratio of the time- and disorder-averaged population, with $\phi_C = 0.64\pi$. **(c)** The localization length, determined by an exponential fit of the time- and disorder-averaged A sublattice site populations versus distance, with $\phi_C = 0.68\pi$. **(d)** The conductivity, determined based on the time- and disorder-averaged probability of the particle occupying the rightmost and leftmost sites, with $\phi_C = 0.69\pi$. In all plots, experimental data is presented in the left subpanel and results from numerical simulations of an ideal model are presented in the right subpanel.

# S6. EXTENDED SIMULATION RESULTS

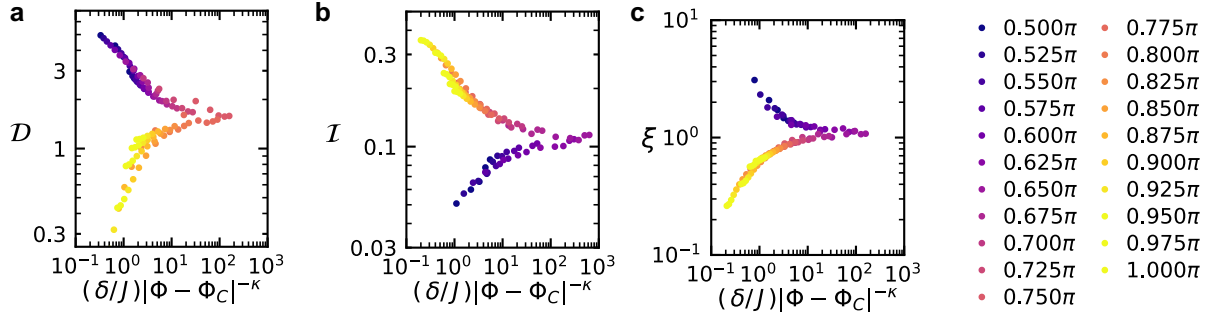

FIG. S14. **Simulated single-parameter scaling for a 12-unit-cell lattice** with uncorrelated disorder. Dynamics are simulated for time $t = 30/J$. **(a)** The RMS position of the time- and disorder-averaged population, with $\phi_C = 0.76\pi$ and $\kappa = 1.4$. **(b)** The inverse participation ratio of the time- and disorder-averaged population, with $\phi_C = 0.64\pi$ and $\kappa = 1.8$. **(c)** The localization length, determined by an exponential fit of the time- and disorder-averaged A sublattice site populations versus distance, with $\phi_C = 0.64\pi$ and $\kappa = 1.4$.

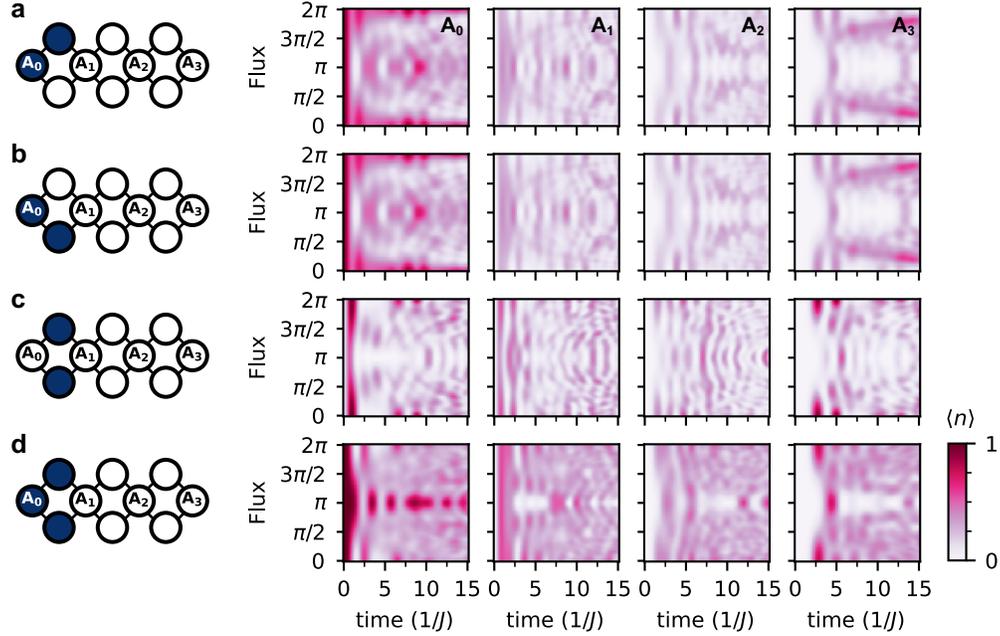

FIG. S15. **Simulated dynamics of multiple interacting particles** in the lattice, accompanying Fig. 5 of the main text. At right, the numerically simulated population at all A sublattice sites as a function of time and flux are shown, following initialization of particles at the sites indicated by dark blue coloring in the diagrams at left. **(a)** Two particles are initialized the leftmost unit cell, one at the A sublattice site and one at the B sublattice site. **(b)** Two particles are initialized in the A and C sublattice sites. **(c)** Two particles are initialized in the B and C sublattice sites. **(d)** Three particles are initialized in the A, B and C sublattice sites.



\end{document}